\documentclass[traditabstract]{aa}  
\usepackage{graphicx,amsmath}

\newcommand{\spitzer}{\emph{Spitzer}}

\begin{document}

\title{Search for rings and satellites around the exoplanet CoRoT-9b using Spitzer photometry}
\author{A. Lecavelier des Etangs\inst{1} 
                \and G.~H\'ebrard\inst{1,2}     
                \and S.~Blandin\inst{1}
                \and J.~Cassier\inst{1}
                \and H.\,J.~Deeg\inst{3,4}
                \and A.\,S.~Bonomo\inst{5}
                \and F.~Bouchy \inst{6}
                \and J.-M.~D\'esert\inst{7} 
                \and D.~Ehrenreich\inst{6} 
                \and M.~Deleuil \inst{8}
                \and R.\,F.~D\'{\i}az\inst{9,10}
                \and C.~Moutou\inst{11}
    \and A.~Vidal-Madjar\inst{1}}
   
\titlerunning{Search for rings and satellite around the exoplanet CoRoT-9b}

\offprints{A.L. (\email{lecaveli@iap.fr})}

   \institute{
   Institut d'astrophysique de Paris, CNRS, UMR~7095 \& Sorbonne Universit\'es, 
   UPMC Paris 6, 98 bis bd Arago, 75014 Paris, France
\and
   Observatoire de Haute-Provence, CNRS/OAMP, 04870 Saint-Michel-l'Observatoire, France
\and
   Instituto de Astrof\'{\i}sica de Canarias, E-38205 La Laguna, Tenerife, Spain
\and
   Universidad de La Laguna, Dept. de Astrof\'\i sica, E-38206 La Laguna, Tenerife, Spain   
\and
         INAF - Osservatorio Astrofisico di Torino, via Osservatorio 20, 10025 Pino Torinese, Italy
\and
   Observatoire de l'Universit\'e de Gen\`eve, 51 chemin des Maillettes, 1290, Sauverny, Switzerland 
\and
   Anton Pannekoek Institute for Astronomy, University of Amsterdam, 1090 GE Amsterdam, The Netherlands
\and
   Aix Marseille Universit\'e, CNRS, LAM (Laboratoire d'Astrophysique de Marseille) UMR 7326, 13388 Marseille, France
\and
   Universidad de Buenos Aires, Facultad de Ciencias Exactas y Naturales, Buenos Aires, Argentina
\and
   CONICET - Universidad de Buenos Aires, Instituto de Astronom\'ia y F\'isica del Espacio (IAFE), Buenos Aires, Argentina
\and
   Canada-France-Hawaii Telescope Corporation, CNRS, 65-1238 Mamalahoa Hwy, Kamuela, HI 96743, USA
}   
   \date{} 
 
  \abstract{
Using \emph{Spitzer} photometry at 4.5 microns, we search for rings and satellites around the
long period transiting planet CoRoT-9b. We observed two transits in 2010 and 2011. 
From their non-detection, we derive upper limits 
on the plausible physical characteristics of these objects in the planet environment. 
We show that a satellite larger than about 2.5~Earth radii is excluded at 3-$\sigma$ 
for a wide range of elongations at the two epochs of observations. 
Combining the two observations, we conclude that rings are excluded for a wide variety 
of sizes and inclination. We find that for a ring extending up to the Roche limit, its
inclination angle from the edge-on configuration as seen from the Earth must be lower than 13$\degr$ 
in the case of silicate composition and lower than 3$\degr$ in the case of material with water ice density. 

%
}


\maketitle
%

\section{Introduction}

Satellites and rings are common features in the solar system. 
All the solar system planets apart from Mercury and Venus host satellites;
more than 80~satellites with diameters larger than 10\,km are known today, 
including the six largest ones with diameters between 3000 and 5300\,km.
Some dwarf planets (Pluto, Eris, and Haumea) also host satellites. 
Rings are also present around the four giant planets Jupiter, Saturn, Uranus, 
and Neptune; recently, narrow and dense rings have been discovered around Chariklo, 
a small Centaur object beyond Saturn. Further, the \emph{Cassini} 
spacecraft revealed evidence for the possible past presence of rings around the 
Saturnian satellites Rhea and Iapetus (Sicardy et al.\ 2017).

The detection of thousands of exoplanets since the mid 90's 
has triggered the search for extrasolar satellites and rings around them. 
Exomoons and exorings are expected to allow constraints to be put 
on small body formation and evolution processes in planetary systems. Many models 
describe the formation of solar system satellites (e.g., Ogihara \& Ida~\cite{ogihara12}); 
some implying impacts or captures (e.g., Agnor et al.~\cite{agnor04},~\cite{agnor06}; see also 
Barr et al.~\cite{Barr et al.2017}, and Barr~\cite{Barr2017}). 
Satellites are also found to play a key role in the evolution of the orbital parameters of some 
planets, as in the case of Uranus (Morbidelli et al.~\cite{morbidelli12}).
In addition to allowing similar studies for exoplanetary systems, exomoons could also provoke interest in terms of potential habitability (e.g., Williams et al.~\cite{williams97}; Heller~\cite{heller12}; 
Heller et al.~\cite{Heller2014}).
Indeed, whereas only a few habitable rocky planets are known, numerous giant exoplanets are known 
and characterized in the habitable zone of their host stars (D\'iaz et al.~\cite{Diaz2016}). 
The detection of a satellite around 
one of them would have a particularly high impact. In addition to the question of habitable moons,
the presence of satellites could also play a role in the habitability of their planets. A famous case is 
the Moon, which is known to stabilize the Earth's obliquity and thus helps to avoid dramatic climate changes, 
which impact the evolution of life (Laskar et al.~\cite{laskar93}).
Finally, the detection of satellites and rings could be a way to probe the nature of 
transiting planets, their formation, and evolution history.
So the detection of exomoons and exorings would be of high interest.

Several detection methods have been proposed. 
All are challenging as the expected signals are extremely small.
The first of them is the transit method in photometry, which is well adapted to searching for material in 
the close environment of transiting planets. Transits of satellites and rings would present clear 
signatures in light curves 
(e.g., Sartoretti \& Schneider~\cite{sartoretti99}; Barnes \& Fortney~\cite{Barnes2004}; 
Arnold~\cite{arnold05}; Ohta et al.~\cite{otha09}; Heller~\cite{Heller2017}). 
More elaborated methods based on transits could improve the sensitivity for exomoon detection; they 
include variation scatter peak (Simon et al.~\cite{simon12}) or the orbital sampling effect (Heller~\cite{heller14}).
Still based on transits, dynamical effects could also reveal the presence of satellites; they would produce 
TTV (transit timing variations: Simon et al.~\cite{simon07}; Lewis et al.~\cite{lewis08}; Kipping~\cite{kipping09a};
Lewis~\cite{Lewis2013}) and/or 
TDV (transit duration variations: Kipping~\cite{kipping09a},~\cite{kipping09b}; Awiphan \& Kerins~\cite{awiphan13}).
Other proposed methods to detect exomoons include 
microlensing (Han \& Han~\cite{han02}; Bennett et al.~\cite{bennett14}), 
spectroscopy (Simon et al.~\cite{simon10};  Kaltenegger~\cite{kaltenegger10}), and even
direct imaging (Agol et al.~\cite{agol15}). Reflected light and spectroscopy have 
also been proposed in the search for exorings (Arnold~\& Schneider~\cite{arnold04}; Santos et al.~\cite{Santos2015}).

Despite several attempts, neither exomoon nor exoring detections have been clearly established up to now.
One of the deepest searches for satellites was presented by Kipping et al.~(\cite{kipping15}) who
reported no compelling detection in a survey of tens of systems using Kepler data and a photodynamical model
using both the transit method and dynamical effects. 
Bennett et al.~(\cite{bennett14}) reported the detection of the microlensing event MOA-2011-BLG-262Lb,
which could be due to a sub-Earth-mass satellite orbiting a free-floating gas giant planet, but another scenario 
with no satellite cannot be excluded.
A deep search for exorings surrounding 21 planets, mostly hot-jupiters, using Kepler photometry 
also yields non-detection (Heising et al.~\cite{Heising2015}).  
Solid surrounding material has been proposed around some exoplanets, including 
$\beta$~Pictoris\,b (Lecavelier des Etangs et al.~\cite{lecavelier95},~\cite{lecavelier16}) and
Fomalhaut\,b (Kalas et al.~\cite{kalas13}). The case of 1SWASP\,J140747.93-394542.6 
is particularly rich but uncertain, as the possible ring around the unseen planet J1407b could be sculpted by the 
presence of a satellite (Kenworthy \& Mamajek~\cite{kenworthy15}). 
So, whereas the existence of exomoons is more than likely, their existence is not demonstrated up to now. 
Their occurrence rate and their properties remain unknown.

CoRoT-9b (Deeg et al.~\cite{deeg10}) is a particularly favorable case for satellite and ring exploration. 
Indeed, it was the first discovered planet to present all the properties of being 
transiting, giant, and far enough from 
its host star to have an extended sphere of gravitational influence (Hill sphere).
CoRoT-9b is a 0.84~Jupiter-mass planet orbiting a G3 main sequence star 
on a nearly circular orbit with a semi-major axis of 0.402~au (a Mercury-like orbit). 
The orbital period is 95.3~days, and the impact parameter of the transits is almost zero, 
resulting in an 8-hour transit duration. 
To have stable orbits, the rings and satellites must be included well within
the Hill sphere where the gravity of the planet dominates the gravity of the star. The stable
prograde orbits are typically within about 0.4~times the radius of the Hill sphere 
(Hinse et al.~\cite{hinse10} and reference therein).   
The size of the Hill sphere is proportional to the periastron distance
(Lecavelier des Etangs et al.~\cite{lecavelier95}). 
Most of the transiting planets known before CoRoT-9b were hot-Jupiter types, 
or pass close to their star at the short periastron distance of their orbit, like HD\,80606b, 
causing them to be strongly irradiated and to have confined Hill spheres. 
For instance, the radius of the Hill sphere around HD\,209458b is about $0.4\times 10^6$\,km, 
or four planetary radii. 
In contrast, CoRoT-9b is a much cooler planet, with a periastron not shorter 
than 0.33~au (Bonomo et al.\ 2017).  As a consequence, CoRoT-9b has an extended Hill sphere 
$R_{\rm Hill} \approx a (1-e) \sqrt[3]{\frac{M_p}{3 M_\star}}  \approx 3.5 \times 10^{6}  $\,km, 
where $a$ and $e$ are the semi-major axis and the eccentricity of its orbit
and $M_p$ and $M_\star$ the planetary and stellar masses, respectively.
This Hill sphere radius is two to three times larger than the
orbit of the major Jupiter and Saturn satellites like Callisto and Titan, which have orbital
distances of 1.8 and $1.2\times 10^6$\,km, respectively.
We highlight also that all satellites of Jupiter and Saturn, including minor satellites, are within 
half of the Hill sphere radius of their planet. 
By analogy, CoRoT-9b presents good prospects for having rings and satellites 
in the inner part of its Hill sphere.
Moreover, the brightness of the CoRoT-9b host star ($V=13.7$) is high enough to allow 
accurate photometry. Consequently, this giant planet is one of the most favorable cases 
to search for transiting solid material such as rings and satellites in its environment. 

With that in mind we observed photometric transits of CoRoT-9b with the \spitzer\ space 
observatory using the IRAC camera at 4.5\,$\mu$m on the post-cryogenic mission.
A sufficiently large satellite around CoRoT-9b would have a clear signature in the transit light curve 
with a duration similar to the planetary transit, that is, about 8~hours. A 2-R$_{\rm Earth}$ satellite 
would produce a transit depth of $\sim 4 \times 10^{-4}$. Rings would have signatures in the 
shape of the light curve during ingress and egress, with a transit depth that could be up to $10^{-3}$.
Thanks to its high sensitivity and its Earth-trailing heliocentric orbit allowing long-duration 
continuous observations, \spitzer\  is particularly well adapted for the search of tiny 
signals in a light curve of a planetary transit.

Observations have been carried out in 2010 and in 2011. These observations are 
presented in Sect.~\ref{Observations}, and the data analysis is described in Sect.~\ref{Data analysis}.
The search for satellites and rings are presented in Sects.~\ref{Search for satellites}
and~\ref{Search for rings} together with the derived upper limits, before concluding in 
Sect.~\ref{Conclusion}.

\section{Observations}
\label{Observations}

The long orbital period of the planet and the observability windows make 
CoRoT-9b transits observable with \spitzer\ particularly rare. 
Only two transits were observable between 2010 and 2013; we observed both of them.

We first observed\footnote{DDT program \#546.} the transit centered on June 18, 2010 at 01h (UT) 
with the IRAC camera of \spitzer\  (Fazio et al.~\cite{fazio04}). Only two infrared channels of the IRAC 
camera are available in the post-cryogenic \spitzer. We chose to observe with only one of the two 
channels in order to avoid repointing the telescope during the transit, thus reducing overheads. 
This allows the target to be located on the same part of the detector during all the observation 
sequence and thus to reduce systematic effects due to imperfect flat-field corrections and 
intra-pixel sensitivity variations. We chose the Channel~2 at 4.5~$\mu$m for the 
observation since it has the best noise properties. This wavelength also has a lower limb-darkening 
effect than Channel~1 at 3.6~$\mu$m. This, as well as the lower uncertainty on the limb darkening effect, favors the search for ring signatures in the 
ingress and egress light curves. 
The observations were executed using IRAC's stellar mode. 
While planning the observation in the Astronomical Observing
Request (AOR) format, we carefully selected a pixel area
avoiding dead pixels. We also purposely did not dither the pointing
in order to keep the source on a given pixel of the detector, and increase 
the photometric accuracy. This common observational
strategy matched that of our previous \emph{Spitzer} observations
of HD189733 (Ehrenreich et al.\ 2007; D\'esert et al.\ 2009, 2011a).

The run\,\#1 observation was secured between  June 17 at 10\,h and June 18 at
15\,h (UT), 2010. We acquired 3338 consecutive images, each of them obtained in 
whole array with a 30-sec integration time (26.8-sec effective integration time per pixel).
Such exposure time clearly avoids saturation of the pixels.  
The 29-hour total duration of the sequence covers the transit of the half inner part of the 
planetary Hill sphere where the rings and large satellites are supposed to be located. 
This includes 10.5~hours before and 10.5~hours after  the transit to cover 
$\sim1.7\times10^6$\,km around the planet.

The second transit observed during run\,\#2 was centered on 
July 04, 2011 at 03h (UT)\footnote{GO7 program \#70031.}. 
We acquired 4376 consecutive images between July 03 
at 8\,h and July 04 at 21\,h (UT), with the same IRAC setting as for the first transit.
This provides more than 37 hours' coverage corresponding to $\sim2.2\times10^6$\,km
around the~planet. The longer duration of this second observation run 
was chosen to improve the coverage of the out-of transit light curve and 
increase the chance of detecting a putative satellite at large distance from the planet: 
the ingress/egress of the satellite at large elongation may require the observation at about 
$\sim$15 hours
before/after the central time of the transit of the planet (see Sect.~\ref{Search for satellites}).

\section{Data analysis}
\label{Data analysis}

\subsection{\emph{Spitzer} photometry}

The analysis was done using the \emph{Spitzer}/IRAC Basic Calibrated Data (BCD)
of the 7714~frames. These frames are produced by
the standard IRAC calibration pipeline and include corrections
for dark current, detector nonlinearity, flat fielding, and conversion
to flux units. To extract the light curves, we first find the center of the point-spread
function (PSF) of the star to a precision of 0.01 pixels by 
computing the weighted centroid of the star. 
To obtain the photometric measurement, we then used the APER routine of 
the IDL Astronomy Library\footnote{http://idlastro.gsfc.nasa.gov/homepage.html}
to perform a weighted photometry within an aperture of a given radius (Horne 1986;
Naylor 1998). After various tests, we decided to use a radius of 2.3 pixels, which  
minimizes the final rms of the light curves. 
The background level for each image was determined with APER by
the median value of the pixels inside an annulus, centered on
the star, with an inner and outer radius of 10 and 15 pixels, respectively.
The PSF, used for weighting, is estimated as the median of the background-subtracted fluxes. 
The estimated error on the weighted integrated flux is calculated as the square root
of the photon-noise quadratic sum (Horne 1986; Naylor 1998).
After producing a time series for each observation, 
we iteratively select and trim outliers 
by comparing the measurements to a transit light curve model. 
Doing so, we remove any remaining points affected by transient phenomena. 
At the end, we discarded 170~exposures in run\,\#1 
and 522~in run\,\#2 ; 
we ended up with a total of 3168~photometric measurements in run\,\#1 
and 3854 in run\,\#2.
The decrease of the measured rms when binning the measurements up to 100 pixels showed 
that the red noise is negligible at least on a timescale of one hour. 

For the 2010 data, we considered the time given in barycentric modified 
Julian Date (BMJD), from which we removed a constant value of 55364. 
This yields a zero reference time on June 17, 2010 at 0h00 UT.
For the 2011 data, we removed a constant value of 55745, 
which yields a zero reference time on July 3, 2011 at 0h00 UT.

\subsection{Correction of systematics}

The \spitzer/IRAC photometry is known to be systematically affected by
the so-called \textit{pixel-phase effect}. This effect produces an oscillation of the
measured fluxes due to the \spitzer\ telescope jitter and the intra-pixel
sensitivity variations on the IRAC detector (see e.g., Charbonneau et al.~\cite{Charbonneau2005};
D\'esert et al.~\cite{Desert2009, Desert11b}).
To correct for these systematics, 
we used
the method developed by Ballard et al.\ (2010). A map of the sensitivity function
is calculated using the residuals of the measurements in a 
non-corrected light curve compared to a theoretical light curve in the 
neighborhood of each pixel (see Eq.~1 in Ballard et al.\ 2010). 
For each measurement, we obtained a weighted-sensitivity correction factor $W$, 
which is used to correct the measurements as a function of the barycenter position of the target. 
The typical length scale of the Gaussian used to smooth the sensitivity map is 
characterized by $\sigma_x$ and $\sigma_y$ in the $x$ and $y$ direction. 
The values of $\sigma_x$ and $\sigma_y$ are obtained by minimizing the {\it rms} 
in the final light curve. 

We also implemented the BLISS algorithm, which uses a bilinear interpolation on a grid to 
map the sensitivity variation as a function of the position of the target on the detector 
(Stevenson et al.\ 2012). We concluded that the BLISS algorithm does not provide
significant improvement compared to the Ballard et al. algorithm: we found that the improvement 
in the rms vanished if, for each corrected measurement, the residual of that measurement 
was excluded from the map calculation. We therefore decided to use the Ballard et al. algorithm
for all our photometric measurement extractions. 

\subsection{Light curves}

The final light curves for run\,\#1 (2010) and run\,\#2 (2011) observations are plotted in 
Fig.~\ref{Light_Curve_2010} and~\ref{Light_Curve_2011}. The flux measurements are fitted
using a theoretical transit light curve given by the equations of Mandell \& Agol (2002) 
in the small planet approximation. 
The limb darkening coefficients for the bandpass of the \emph{Spitzer} Channel~2 at 4.5$\mu$m 
are obtained from the table of Sing (2010). 
For a star with $T=5630K$, $[Fe/H]=0$, and $\log g=4.5$ and
fixing $c_1=0$, interpolation of Sing's Table gives $c2= 0.8071$, $c3=-0.9785$, and 
$c4=0.3973$.



A first fit reveals a bump in the light curve of run\,\#1 (Fig.~\ref{Light_Curve_2010}). 
It could be due to an instrumental artifact or to the planet transiting in front of a stellar spot, however we make no conclusions as 
 CoRoT-9 is not known to be particularly active. 
The bump is seen from $t=24$\,h to $t=25.5$\,h, and hereafter the corresponding data have 
been excluded for the analysis presented here. 
  
The fits to the light curves have six free parameters: 
$b$, the impact parameter in units of stellar radius, $T_0$, the central time of the transit, 
$R_p/R_*$, the planet to star radius ratio, $v$ the tangential transit velocity in stellar radius per unit of time, 
and $A_{0,1}$ the two parameters of the first degree polynomial fitting the light curve baseline. 
The tangential transit velocity $v$ is related to the semi-major axis $a$ by the equation 
$v/R_* = (1+e\cos \varpi)/(\sqrt{1-e^2})\times 2\pi /P \times a/R_*$, 
where $e$, $\varpi$, and $P$ are the eccentricity, the longitude of periastron, 
and the period of the orbit, respectively. 
The eccentricity is not a parameter of the model because
it does not affect the shape of the light curve (see H\'ebrard et al.\ 2010) ; 
it is only needed to convert the measured $v/R_*$ into an estimate of $a/R_*$.  
We also let the weighted-sensitivity $W$ free to vary in calculating the parameter and its error bar estimates.  
This ensures that the error bars are not underestimated by a bias introduced by a miscalculation of $W$. 

We found that these light curves are well fitted by $b=0.00\pm0.25$, 
$T_0$$_{run\#1}=24.645\pm 0.020$\,h, $T_0$$_{run\#2}=26.795\pm 0.025$\,h,
$v/R_*=0.271_{-0.010}^{+0.002}$\,h$^{-1}$. 
We found a slightly different $R_p/R_*$ radius ratio in 
the two light curves : 
$R_p/R_*$$_{run\#1}=0.1167\pm0.0011$ and $R_p/R_*$$_{run\#2}=0.1140\pm0.0015$. 
This small difference (1.5-$\sigma$) could be linked to the bump seen in run\,\#1.
If the bump is due to an occulted spot, this would yield a higher relative surface brightness 
of the spot-free transited part of the stellar disk used in the fit 
(we recall that we excluded from the fit the measurements 
obtained during the bump, see Fig.~\ref{Light_Curve_2010}) 
and therefore a bias towards a higher transit depth.

The determination of refined parameters for the CoRoT-9 planetary system is beyond the scope 
of the present paper. This is done by Bonomo et al.\ (2017), 
who use the present {\emph{Spitzer} data as well as other datasets. 
  
\begin{figure*}[p]
\begin{center}
\includegraphics[angle=0,width=0.85\textwidth]
{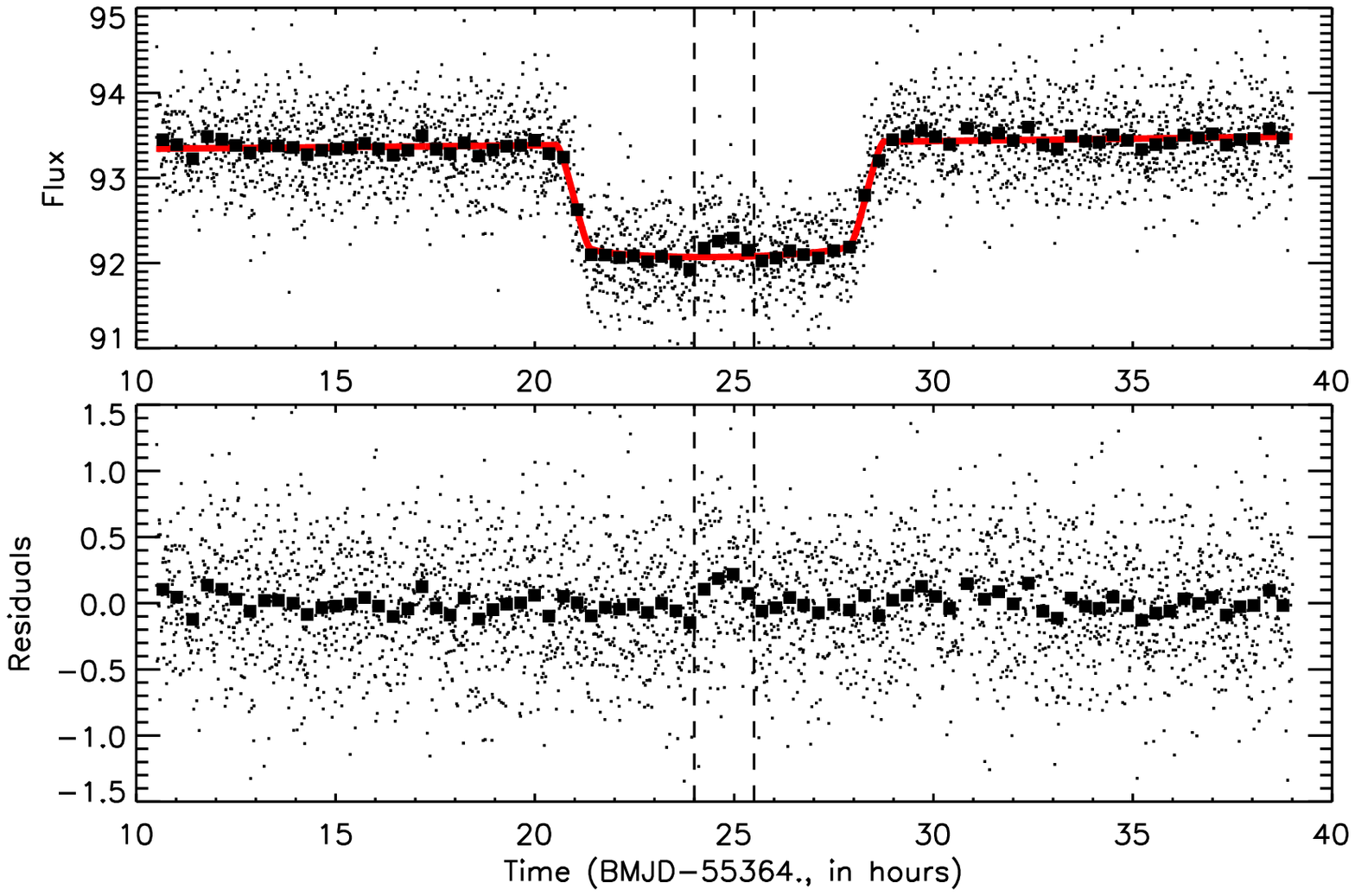}
\caption[]{Light curve of CoRoT-9 during the transit of CoRoT-9b on June 18, 2010. 
Each dot represents an individual photometric measurement. 
Large squares show the data measurements rebinned by 40. 
The thick red line shows the best fit with an exoplanet transit model. 
The bottom panel gives the residual of the measurements after subtraction of the best fit model. 
The two vertical dashed lines represent the time range of the estimated time of the photometric bump 
that could be due to the transit in front of a stellar spot. 
The data taken within this time range have been excluded from the analysis.  
\label{Light_Curve_2010}}
\end{center}
\end{figure*}

\begin{figure*}[h]
\begin{center}
\includegraphics[angle=0,width=0.85\textwidth]
{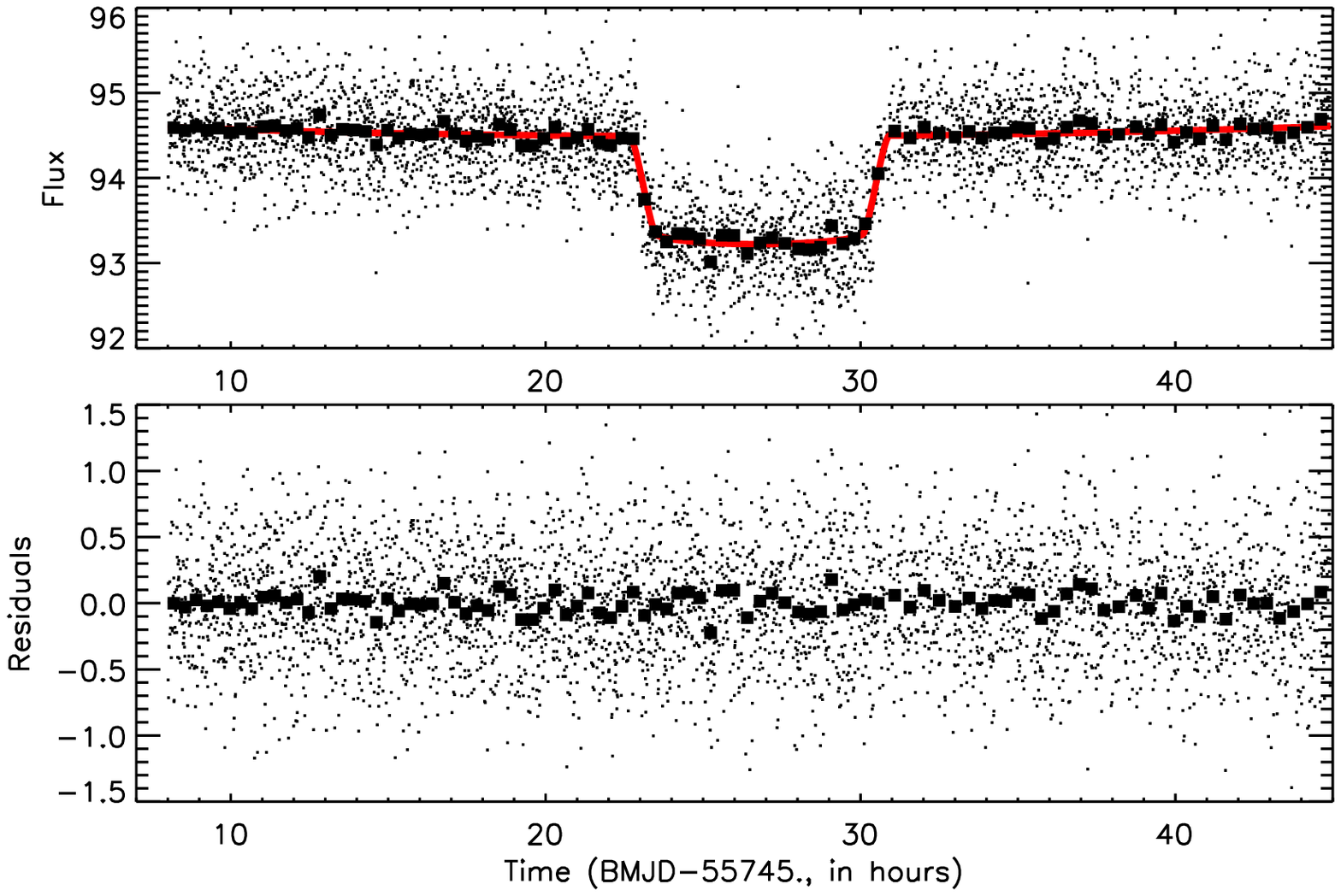}
\caption[]{Same as Fig.~\ref{Light_Curve_2010} for the light curve of CoRoT-9 during the transit of CoRoT-9b on July 4, 2011.
\label{Light_Curve_2011}}
\end{center}
\end{figure*}

\section{Search for satellites}
\label{Search for satellites}

\subsection{Method}

In all of the following, we search in the light curves for the signature of the transit 
of an additional body in front of the star, which could be a satellite. 
If a satellite is present in the environment of the planet, it will produce an additional occultation 
when it transits the star. The satellite transit depth depends on the satellite to star radius ratio $R_{sat}/R_*$. 
The transit epochs or central time of the satellite ($T_{0s}$) depend on the elongation of the satellite 
relative to the planet projected on the sky and on the direction of the planet's trajectory.   

To perform the search for the satellite, we fitted the light curves of 2010 and 2011 observations 
with theoretical light curves including a satellite with the two additional fixed parameters defined above
($R_{sat}/R_*$ and $T_{0s}$) whilst keeping the other parameters describing the planet and its orbit free to vary. 
For each satellite configuration, we calculated the resulting $\chi^2(R_{sat}/R_*,T_{0s})$ 
and its difference with the best $\chi^2_0$ of the fit without a satellite: 
$\Delta\chi^2=\chi^2(R_{sat}/R_*,T_{0s})-\chi^2_0$.  
Although the light curve noise may not be exactly Gaussian, this $\Delta\chi^2$ can be used to quantify 
the improvement in the fit when introducing a satellite in the model. A more rigorous treatment 
would require modeling the data covariance matrix, which is also beyond the scope of this paper. 
Furthermore, we estimate that the effect on our results would be small.
Moreover, to see if the possible improvement of the $\chi^2$ (when $\Delta\chi^2<0$) 
could be due to features in the light curve that are not related to a satellite 
({\it i.e.,} correlated red-noise on transit duration timescale, 
or simply the decrease of the number of degrees of freedom with the addition of two parameters in the model), 
we also considered the theoretical light curves 
with an increase of the flux with the same shape as the flux 
decrease due to a satellite of a given size and elongation: we parametrize those test light curves 
(``bright-satellite'') with a negative planet size $R_{sat}/R_*<0$.

In these calculations to search for a perturbation in the light curve due to a satellite 
(or a ring in the Sect.\ref{Search for rings}), we also left the sensitivity correction 
factors $W$ free to vary. 
This ensures that if a ring or a satellite is detected it cannot be due to a bias introduced by a 
miscalculation of $W$.

\subsection{Results}

The resulting $\Delta\chi^2$ difference as a function of the satellite size and elongation are plotted 
in Figs.~\ref{Contour_sat_2010} and~\ref{Contour_sat_2011} for the 2010 and 2011 observations, respectively.

In the fit to the 2010 data, an improved $\chi^2$ is found 
with $\Delta \chi^2$$\la$-4 for $R_{sat}/R_*$$\sim$0.026 and $T_{0s}$$\sim$36.5\,h, 
and for $R_{sat}/R_*$$\sim$0.025 and $T_{0s}$ in the range 16\,h-21\,h (Fig.~\ref{Light_curve_sat_2010}). 
However, it can be seen in Fig.~\ref{Contour_sat_2010} that similar $\Delta \chi^2$ 
can be found for negative values of $R_{sat}/R_*$, with even $\Delta \chi^2$$<$-9 
for $R_{sat}/R_*$$\sim$-0.028 and $T_{0s}$$\sim27.6$\,h.
We conclude that in the cases where there is an improvement in the fit by including a satellite 
in the transit model, this improvement is likely due to correlated noise and other features 
in the light curves that are not related to the transit of a satellite in front of the star. 
We note that although Bonomo et al.\ (2017) stated that the correlated noise is practically 
negligible, here we are considering correlated noise on the very long timescale of the transit duration, which 
cannot be quantified within the present data.  

The observations of 2011 yield similar results (Fig.~\ref{Contour_sat_2011}). 
The $\chi^2$ improvement is $\Delta\chi^2$$\sim$-3.4 
at the best for $R_{sat}/R_*$$\sim$0.020 and $T_{0s}$$\sim$20.5\,h (Fig.~\ref{Light_curve_sat_2011}). 
Again this is not a significantly better fit to the data compared to
the fits that include a (fake) ``bright-satellite'' like the one at, for instance,  
$R_{sat}/R_*$$\sim$-0.020 and $T_{0s}$$\sim$35\,h, which yields $\Delta\chi^2$$\sim$-5. 
Therefore, we conclude that this best fit including a satellite is most likely a false positive signature 
for a satellite transit in the light curve.  

Nonetheless, we can derive upper limits on the size of a satellite that can be detected in the present 
data. Using 2010 data (Fig.~\ref{Contour_sat_2010}), we derive an
upper limit at 3-$\sigma$ ($\Delta\chi^2$$>$9) for the radius ratio 
of $R_{sat}/R_*<0.025$ ($R_{sat}<2.5$$R_{\oplus}$) if $T_{0s}$ ranges 
from 8\,h to 12\,h and from 27\,h to 31.5\,h. 
These $T_{0s}$ correspond to projected distances 
to the planet of 3.0$-$2.2$\times$$10^6$\,km (0.9$-$0.6~Hill radius) ahead of the planet 
and 0.4$-$1.2$\times$$10^6$\,km (0.1$-$0.3~Hill radius) after the planet at the epoch of observation. 
For other elongations, correlated noise leading to a fake positive signature significantly increases 
the upper limit, and the transit of the planet itself hides the possible signature of a satellite for 
$|T_{0s}-T_{0p}|$$<$2.0\,h, corresponding to a projected distance lower than 0.35$\times$10$^6$\,km (0.1~Hill radius).

The 2011 data yields approximately similar upper limits: we derive the same upper limit 
of 2.5$R_{\oplus}$ for $T_{0s}$ ranging from 29\,h to 38\,h, that is for a projected distance 
of 0.4$-$2.0$\times$$10^6$\,km (0.1$-$0.6~Hill radius) after the planet at the epoch of observation. 

We did not compute upper limits using the 2010 and 2011 transits simultaneously 
as any potential satellite is expected to have a different position at different epochs.

\begin{figure}[p!]
\begin{center}
\includegraphics[angle=0,width=\columnwidth]
{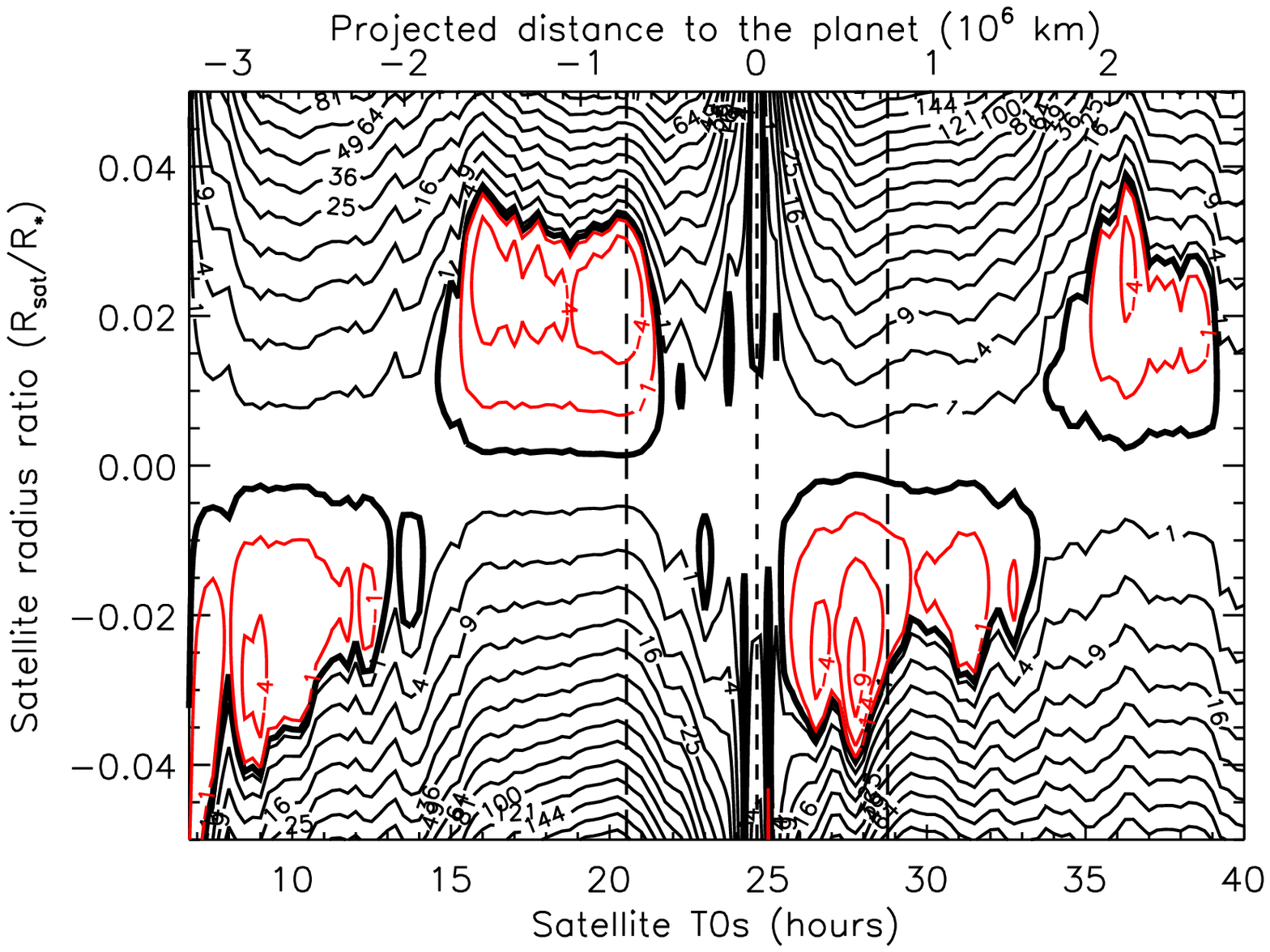}
\caption[]{Difference of the $\chi^2$ between the best fit with a satellite and without a satellite 
as a function of the satellite size and central time of transit, 
measured in the light curve of the 2010 transit. 
The red iso-curves are for negative values of $\Delta\chi^2$, 
where the fit with a satellite is better than the fit without satellite. 
The vertical lines show the central time (short-dashed) and the time of first and fourth contact 
(long-dashed lines) for the transit of the planet. 
The upper x-axis provides the projected distance of the satellite to the planet (elongation) 
for the corresponding central time of the satellite transit ($T_{0s}$). 

\label{Contour_sat_2010}}
\end{center}
\end{figure}

\begin{figure}[htb]
\begin{center}
\includegraphics[angle=0,width=\columnwidth]
{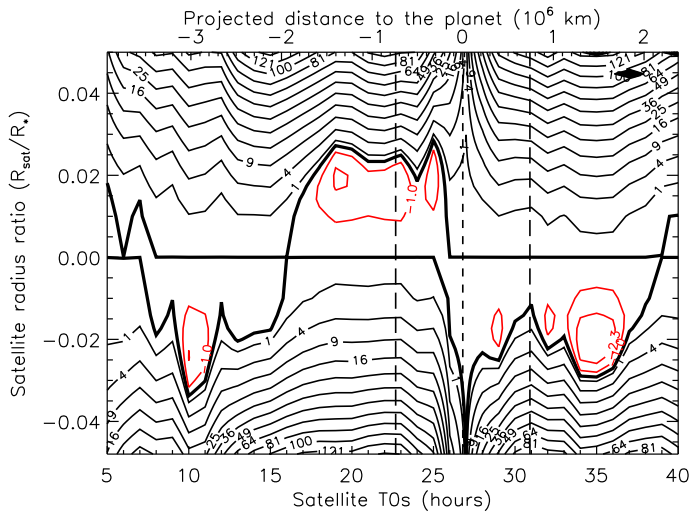}
\caption[]{Same as Fig.~\ref{Contour_sat_2010} using the light curve of the 2011 transit.
\label{Contour_sat_2011}}
\end{center}
\end{figure}

\begin{figure}[htb]
\begin{center}
\includegraphics[angle=0,width=\columnwidth]
{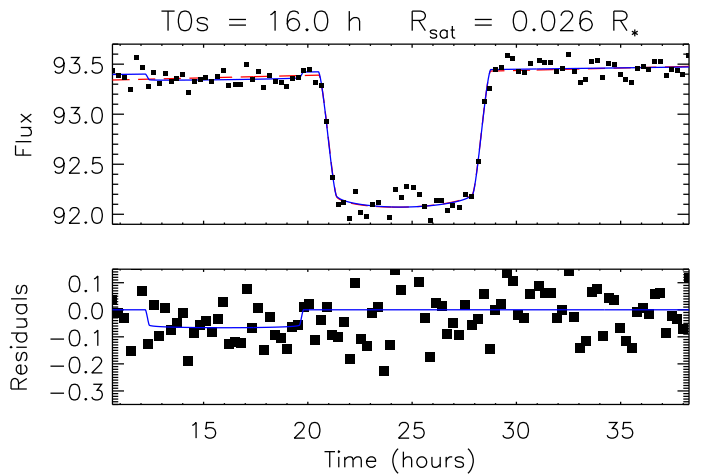}
\caption[]{The light curve of the transit on June 18, 2010, and the fit including a satellite 
with a radius of 0.026 times the planet radius, and with a central time of transit 16.0 hours 
after June 17, 2010, 0h00 UT (blue solid line). 
This fit has a $\chi^2$ of 3162.4, that is smaller by $\sim$6.4 
than the $\chi^2$=3168.8 of the fit without a satellite. 
The data have been rebinned by 31 into 100~individual measurements. 
The red dashed line shows the transit model without a satellite. The model
light curve with the satellite has an additional occultation starting at about 12~hours 
and ending at about 20~hours.
The bottom panel shows the residuals of the data (filled squares) and the light curve of the
model including the satellite, after subtraction of the best fit without a satellite. 
 
\label{Light_curve_sat_2010}}
\end{center}
\end{figure}

\begin{figure}[htb]
\begin{center}
\includegraphics[angle=0,width=\columnwidth]
{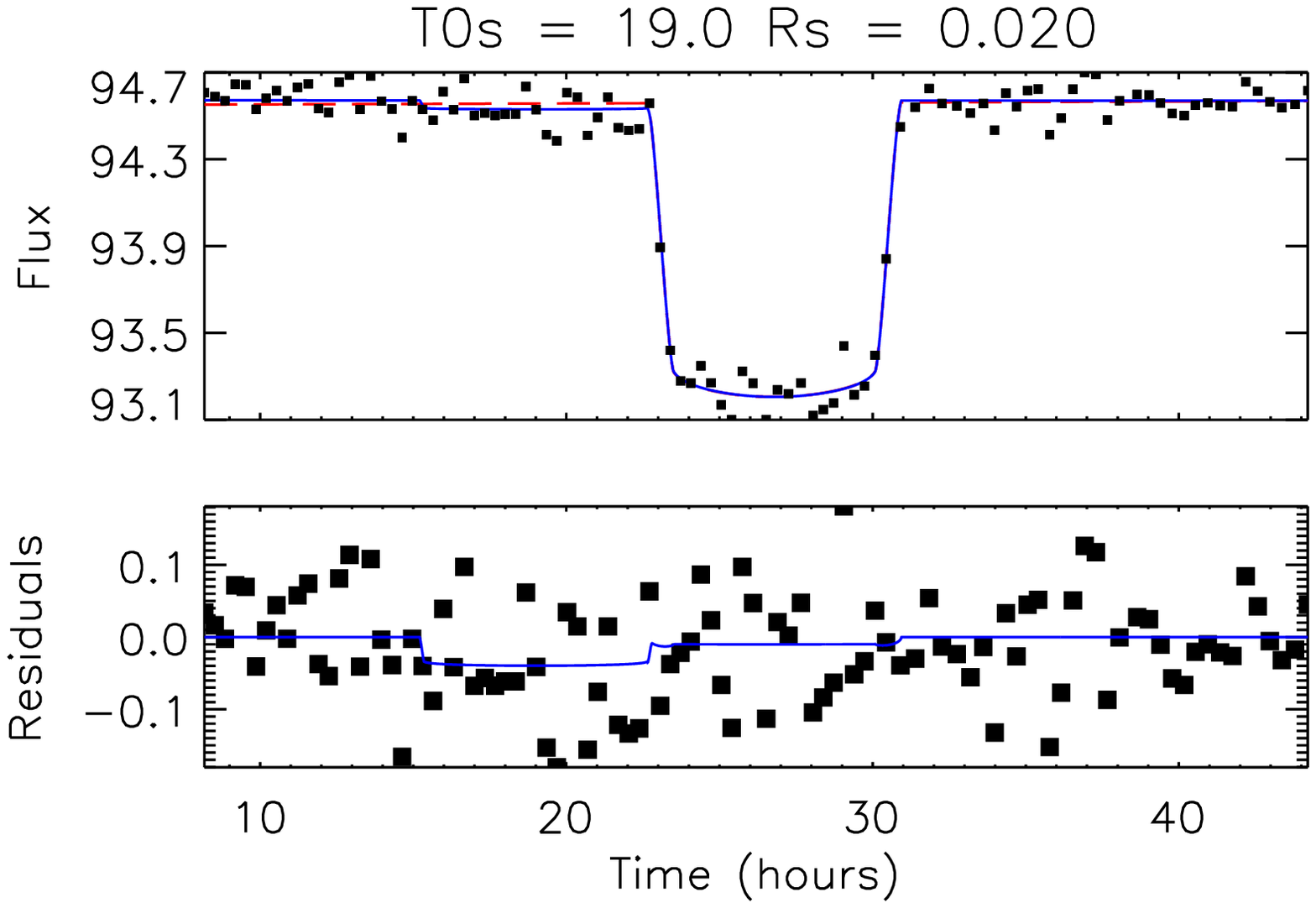}
\caption[]{Same as Fig.~\ref{Light_curve_sat_2010} for the transit of 2011 and 
the fit including a satellite with a radius of 0.020 times the planet radius, 
and with a central time of transit 19.0 hours after July 3, 2011, 0h00 UT. 
This fit has a $\chi^2$ of 3938.8, that is smaller by $\sim$2.5 than the $\chi^2$=3941.3 
of the fit without a satellite. 
Here the additional occultation by the satellite starts at about 15~hours.
\label{Light_curve_sat_2011}}
\end{center}
\end{figure}

\section{Search for rings}
\label{Search for rings}

\subsection{Method}

We also searched in the light curves for the signature of the transit 
of rings surrounding the planet.
If rings are present in the environment of the planet, they affect the shape of the
light curve observed during the ingress and the egress: at first order, the rings, which are 
elongated occulting bodies, make the ingress/egress last longer than a spherical body producing the
same occultation depth. 
To avoid an additional degree of freedom in the considered ring models, we hypothesize 
that the ring has an ascending node close to the plane of the sky 
and thus is oriented such that, as seen from the Earth, 
the main axis of the elliptic ring shadow is in the direction of the planet trajectory. 
For a ring that has a different orientation, at first order, a ring configuration with the same extension 
and thickness would produce a very similar light curve. In the case of a positive detection, 
it would be interesting to explore various inclination configurations in the three-dimensional space that
match the detection. But with no detection, this is not the purpose of the work presented here.    

In the light curve, the ring transit depth and transit duration depend on the apparent thickness and 
size of the elliptic shadow of the ring. The ellipse figure of the ring is characterized 
by its extension from the planet limb ($L_r$) and its apparent thickness or height ($H_r$), 
both expressed in units of stellar radius. From these parameters, we can derive
the semi-major and semi-minor axis of the shadow ellipse occulting the stellar disk to be
$a_r=L_r+R_p$ and $b_r=H_r/2$. 
Therefore, the inclination of the ring axis relative to the plane of the sky, $i_r$, is given by  $H_r/2/(L_r+R_p)=\sin(90\degr-i_r)$.

To perform the search for rings, we fitted the light curves of 2010 and 2011 observations 
with theoretical light curves including a ring with the two additional parameters defined above
($L_r/R_*$ and $H_r/R_*$). 
The area of the rings shadow occulting the stellar disk was calculated using a generalization of 
equations given by Zuluaga et al.~(\cite{Zuluaga2015}) and equations for the
sector and segment areas of an ellipse given by 
Cavalieri's Principle\footnote{Alexander Bogomolny, Area, Sector Area, and Segment Area of an Ellipse, 
from {\it Interactive Mathematics Miscellany and Puzzles}, 
http://www.cut-the-knot.org/Generalization/Cavalieri2.shtml}
to obtain an analytical derivation of 
the area of the intersection of the stellar disk, the planetary disk, and the ring's shadow. 
For each ring configuration, we calculated the resulting $\chi^2(L_r/R_*,H_r/R_*)$ 
of the fit with the given ring 
and the corresponding difference with the best $\chi^2_0$ of the fit without a ring: 
$\Delta\chi^2=\chi^2(L_r/R_*,H_r/R_*)-\chi^2_0$.  
To see if the possible improvement of the $\chi^2$ (when $\Delta\chi^2<0$) 
could be due to features that are not a ring, 
we also considered the theoretical light curves 
with an increase of the flux as we have done in the search for a satellite. We parametrize those test light curves (``bright-ring'') with a negative ring thickness $H_r/R_*<0$.

\subsection{Results}

The resulting $\Delta\chi^2$ difference as a function of the ring size and ring thickness is plotted 
for the 2010 and 2011 observations in Figs.~\ref{Contour_ring_2010} and~\ref{Contour_ring_2011}.
Because the rings are not expected to have changed between the two observations, we also 
produced the same plot obtained by combining the two sets of data (Fig.~\ref{Contour_ring_combine}).
A plot of the light curve and the corresponding fit including a ring is given in Fig.~\ref{Light_curve_ring_2010}. 

None of the fits including rings yield a significantly better fit to the light curves (2010, 2011 or combined) 
than the best fit without a ring. The best improvement in the $\chi^2$ is lower than a unity: 
$\Delta \chi^2$$>$$-1$. The $\Delta \chi^2$ is found to be closely related to the equivalent 
occulting area of the ring, which is proportional to $L_r \times H_r$ (light blue curves 
in Figs.~\ref{Contour_ring_2010}, \ref{Contour_ring_2011} and~\ref{Contour_ring_combine}). 

As for the satellite, despite a non-detection, we can derive upper limits on the size of a ring
that could have been detected in the present data as a function of its thickness. 
Using the combined data set (Fig.~\ref{Contour_ring_combine}), we derive an
upper limit at 3-$\sigma$ ($\Delta\chi^2$$>$9) for the ring size. 
For thickness of $H_r/R_*=0.015$, the 3-$\sigma$ upper limit is $L_r/R_*$$<$0.25; 
for $H_r/R_*=0.030$ and $H_r/R_*=0.050$, the upper limits are $L_r/R_*$$<$0.16 
and $L_r/R_*$$<$0.12, respectively. 

The Roche limit for the ring size in the fluid approximation is about 2.4$R_p$$^3 \sqrt{\rho_p / \rho_r}$, 
where $R_p$ is the planet radius, and $\rho_p$ and $\rho_r$ are the density of the planet 
and the ring material\footnote{We note that rings may be larger than the Roche limit, 
as is known from the presence of Saturn's E ring and the still wider Phoebe ring. 
However, these rings are very tenuous (the denser E Ring has an optical depth of $\la$10$^{-5}$,
Showalter et al.~\cite{Showalter1991}), making the detection of such wide rings 
in transit observations highly unlikely.}, respectively. 
For a ring composed of silicates ($\rho_r$$\sim$2.6\,g\,cm$^{-3}$) or water ices 
($\rho_r$$\sim$ 1\,g\,cm$^{-3}$), we have a Roche limit of 1.7$R_p$ and 2.4$R_p$, respectively. 
Because here the ring size, $L_r$, is counted from the planet limb, these
latter values correspond to the ring sizes of $L_r$=0.7$R_p$=0.08$R_*$ and $L_r$=1.4$R_p$=0.16$R_*$. 
For these ring sizes, the combined data set yields 3-$\sigma$ upper limits on the thickness that are 
$H_r/R_*$$<$0.090 and $H_r/R_*$$<$0.030, respectively (Fig.~\ref{Contour_ring_combine}). 
The ring thickness $H_r$ is related to $i_r$ the ring inclination to the line of sight by 
$H_r/2/(L_r+R_p)=\sin(90\degr-i_r)$; 
we therefore conclude that in the case of a ring extending up to the Roche limit, 
our non-detection constrains its inclination to be $|i_r-90\degr|<13\degr$ in the case of silicates 
and $|i_r-90\degr|<3\degr$ in the case of material with water ice density. 

It is also interesting to know if rings similar to Saturn's and not fully opaque 
could be detected. 
Here forward scattering is negligible because observable signals require scattering surfaces 
that subtend a significant angle as seen from the star, as would be the case for objects on shorter orbital distances.
An example for an observable scattering signal is from KIC12557548 (Rappaport et al.\ 2012), 
an evaporating transiting planet on a 16-hour orbit, but scattering cannot be observable 
for a ring around CoRoT-9b at 0.33~au from the star 
(see also the marginal scattering amplitude compared to the extinction for a comet-like object at 0.3~au 
from the star, in Fig.~4 of Lecavelier et al.\ 1999).
Saturn's inner A-D rings extend to a distance of 137\,000\,km from Saturn's center. 
This corresponds to $L_r$=1.3$R_{\rm Saturn}$, where we use a Saturn equatorial radius of 60\,000\,km. 
For a corresponding value of $L_r$=1.3$R_p$=0.15$R_*$, we have a 3-$\sigma$ upper limit for the thickness
of $H_r/R_*$$<$0.032 or $H_r$$<$0.28$R_p$. Based on profiles of the Saturn rings' optical transmission 
at 0.9~microns (Nicholson et al.~\cite{Nicholson2000}), we use an average optical depth of $\tau$=0.3. 
The transmittance $T_{i}$ of the rings at a given inclination $i_r$ (where 0$\degr$ inclination is for face-on rings) 
is given by the usual equation  $T_{i}$=$e^{(-\tau/\cos i_r)}$. 
For a face-on disk, we obtain a transmittance  $T_{i=0}$=0.75.
For inclined translucent rings, we have an occultation similar to rings with
$H_r$=2$\times(1-T_{i})\times(L_r+R_p)\sin(90\degr-i_r)$. Assuming an optical depth of $\tau$=0.3 
and a ring size $L_r$=1.3$R_p$, from the 3-$\sigma$ 
upper limit of $H_r$$<$0.28$R_p$ we therefore derive that Saturn-like rings with inclination 
as large as 4$\degr$ could still have been detected around CoRoT-9b. 
The chance of missing a Saturn-like ring system is therefore relatively low. 

\begin{figure}[htb]
\begin{center}
\includegraphics[angle=0,width=\columnwidth]
{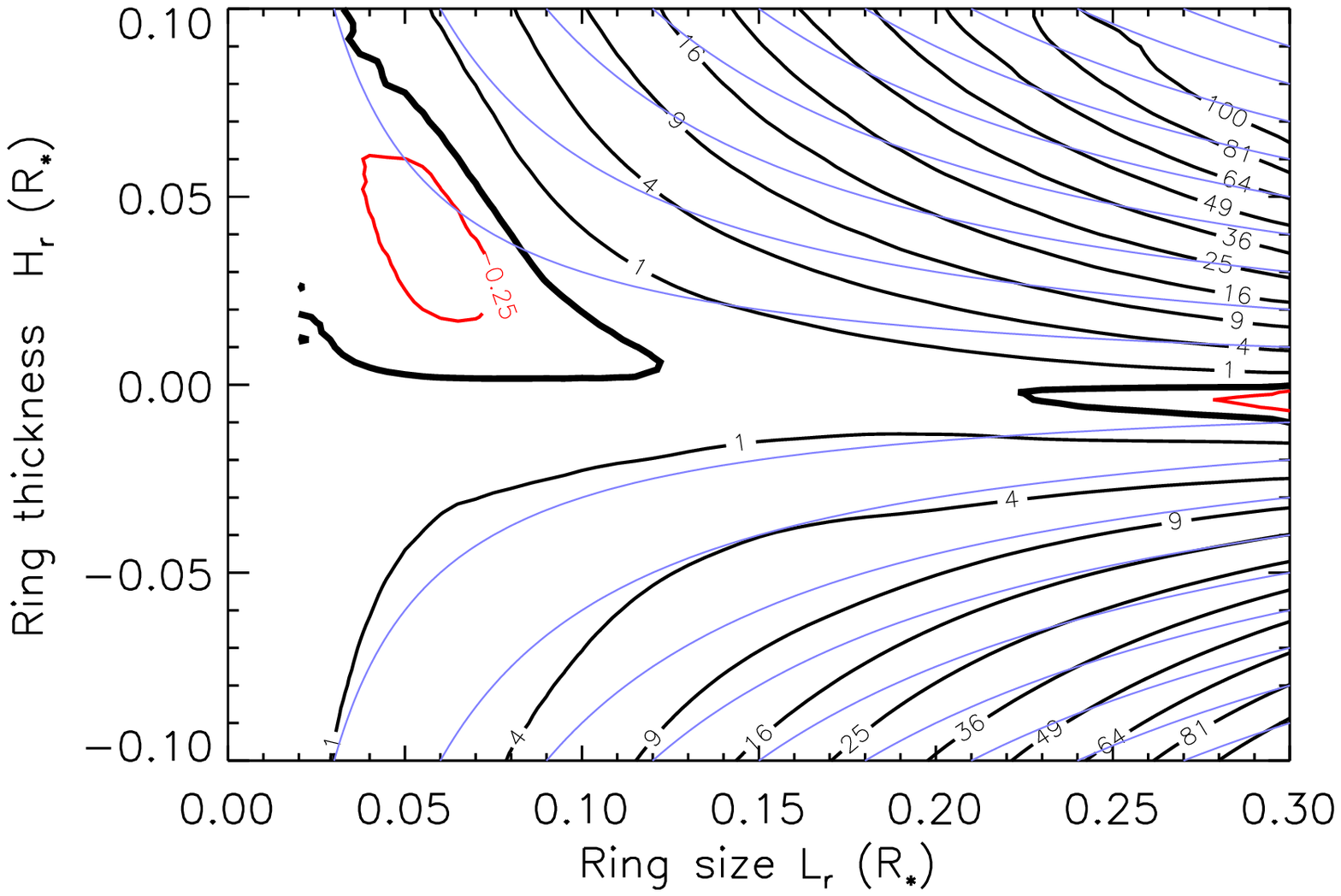}
\caption[]{Difference of the $\chi^2$ between the best fit with a ring and without a ring 
as a function of the ring size and thickness, which is 
obtained using the light curve of the 2010 transit.
Blue thin lines show the iso-curves of the equivalent 
occulting area of the ring $L_r \times H_r$. 

\label{Contour_ring_2010}}
\end{center}
\end{figure}

\begin{figure}[htb]
\begin{center}
\includegraphics[angle=0,width=\columnwidth]
{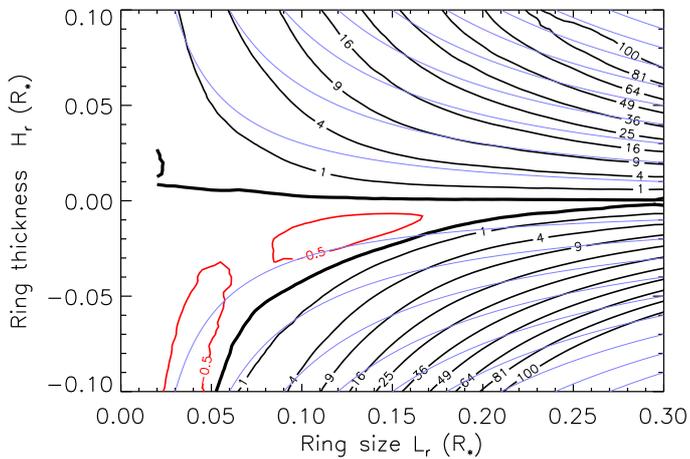}
\caption[]{Same as Fig.~\ref{Contour_ring_2010} for the light curve of the 2011 transit.
\label{Contour_ring_2011}}
\end{center}
\end{figure}

\begin{figure}[htb]
\begin{center}
\includegraphics[angle=0,width=\columnwidth]
{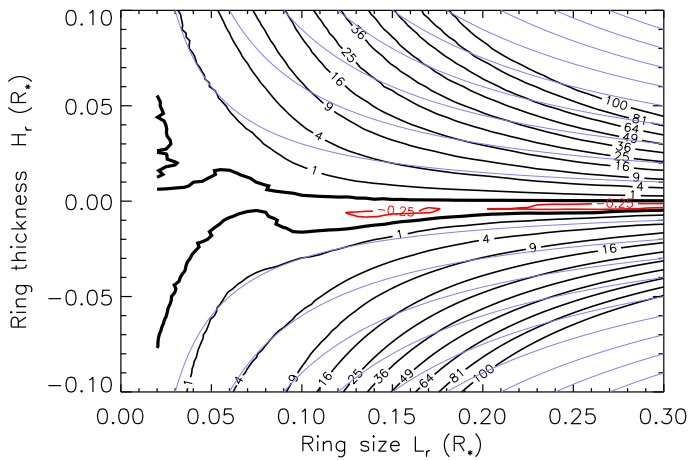}
\caption[]{Same as Fig.~\ref{Contour_ring_2010} using the combination of both the light curves of the 2010 and 2011 transits.
\label{Contour_ring_combine}}
\end{center}
\end{figure}

\begin{figure}[htb]
\begin{center}
\includegraphics[angle=0,width=\columnwidth]
{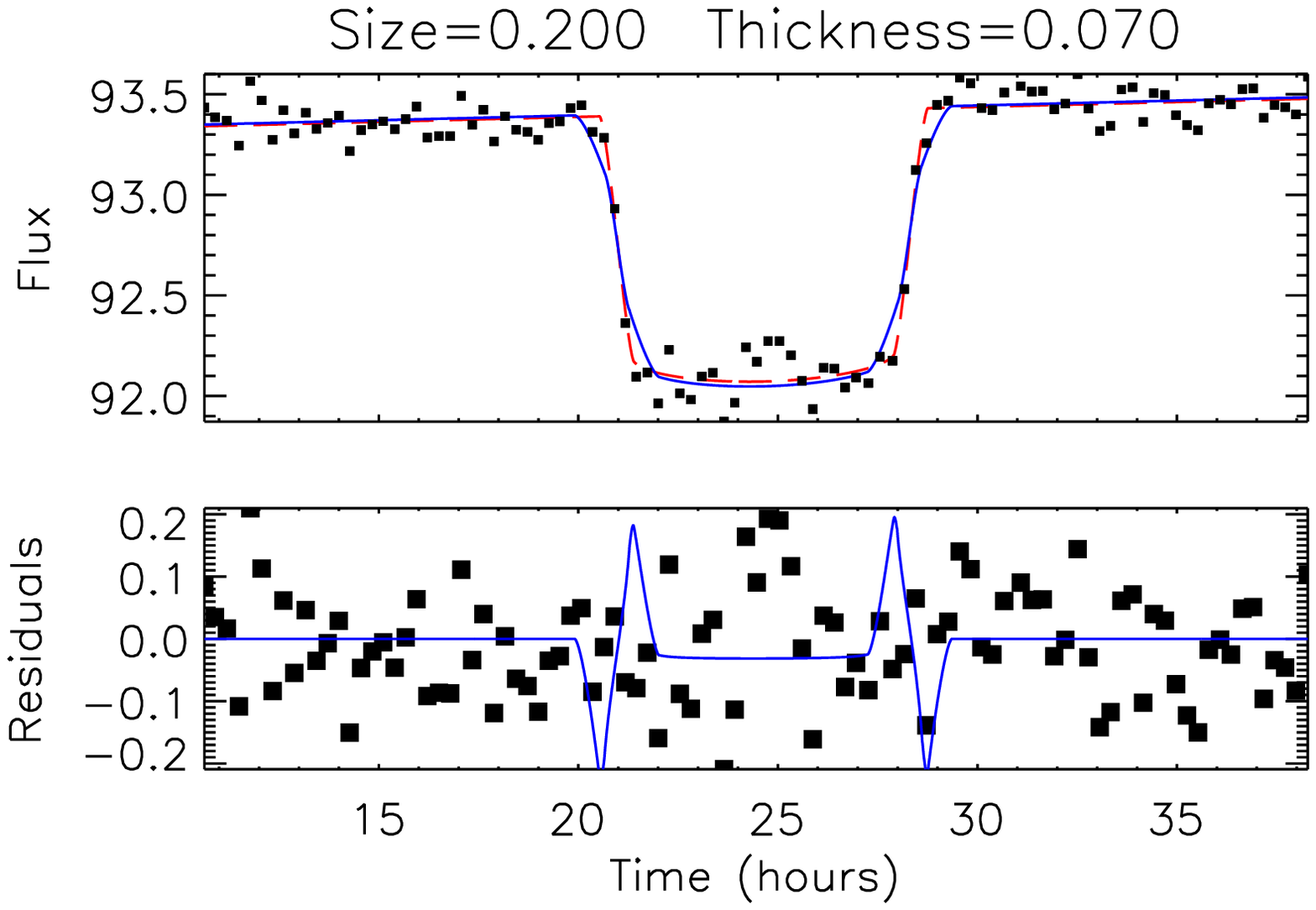}
\caption[]{The light curve of the transit on June 18, 2010, and the fit including a ring 
with a size $L_r$=0.2\,$R_*$ and a thickness $H_r$=0.07\,$R_*$, where $R_*$ is 
the star radius (blue line). 
This fit has a $\chi^2$ of 3204, that is larger by $\sim$35 
than the $\chi^2$ of the best fit without a ring (thin red line). 
Such ring is thus excluded at about 6-$\sigma$ confidence level. 
The bottom panel shows the residuals after subtraction of the best fit without ring.
\label{Light_curve_ring_2010}}
\end{center}
\end{figure}
 
\section{Conclusion}
\label{Conclusion}

Through detailed analysis of two transit light curves of CoRoT-9b, we searched for the signature of a satellite 
and rings in the planet's environment. 
Despite the non-detection, we have been able to derive upper limits for the satellite and the ring sizes. 
The upper limits on the satellite size are not very constraining, nonetheless the presence of satellites
may have been jeopardized by the dynamical instability, which could be at the origin of the planet eccentricity 
(Bonomo et al.\ 2017). 
Since the discovery of CoRoT-9b, a few other exoplanets have been discovered in transit at large orbital distances. 
To enlarge the exploration of the giant exoplanets' environment, these targets deserve to be observed 
at high photometric accuracy. 
In this spirit, it is expected that the forthcoming new space facilities, such as the ESA/Cheops photometric observatory, 
will allow a deeper search of perturbation in the light curves that can be due to small bodies 
in orbit around those planets. 

\begin{acknowledgements}
This work is based on observations made with the \emph{Spitzer} Space Telescope, which is operated by the Jet Propulsion Laboratory, California Institute of Technology under a contract with NASA.
Support for this work was provided by the CNES and by an award from the Fondation Simone et Cino Del Duca. 
We acknowledge the support of the French Agence Nationale
de la Recherche (ANR), under program ANR-12-BS05-0012 ``Exo-Atmos''.
ASB acknowledges funding from the European Union Seventh Framework programme (FP7/2007-2013) under grant agreement No.\ 313014 (ETAEARTH).
HD acknowledges support by grant ESP2015-65712-C5-4-R of the Spanish Secretary of State for R\&D\&i (MINECO).  

\end{acknowledgements}


\begin{thebibliography}{}

\bibitem[2004]{agnor04} 
Agnor, C., Asphaug, E. 2004, \apj, 613, L157

\bibitem[2006]{agnor06} 
Agnor, C. B., Hamilton, D. P. 2006, \nat, 441, 192

\bibitem[2015]{agol15} 
Agol, E., Jansen, T., Lacy, B., Robinson, T. D., Meadows, V. 2015, \apj, 812, 5

\bibitem[2004]{arnold04} 
Arnold, L., Schneider, J. 2004, \aap, 420, 1153

\bibitem[2005]{arnold05} 
Arnold, L. F. A. 2005, \apj, 627, 534

\bibitem[2013]{awiphan13} 
Awiphan, S., Kerins, E. 2013, \mnras, 432, 2549

\bibitem[2010]{Ballard2010}Ballard, S., Charbonneau, D., Deming, D., 
et al.\ 2010, \pasp, 122, 1341 

\bibitem[2004]{Barnes2004}Barnes, J.~W., \& Fortney, J.~J.\ 2004, \apj, 616, 1193 

\bibitem[2017]{Barr2017}Barr, A.~C.\ 2017, arXiv:1701.02125 

\bibitem[2017]{Barr et al.2017}Barr, A.~C., \& Syal, M.~B.\ 2017, \mnras,  

\bibitem[2014]{bennett14} 
Bennett, D. P., Batista, V., Bond, I. A, et al. 2014, \apj, 785, 155

\bibitem[2017]{Bonomo2017}
Bonomo, A. S., H\'ebrard, G., Raymond, S. N., et al.\ 2017, \aap, in preparation

\bibitem[2005]{Charbonneau2005}Charbonneau, D., Allen, L.~E., 
Megeath, S.~T., et al.\ 2005, \apj, 626, 523 

\bibitem[2010]{deeg10} 
Deeg, H.~J., Moutou, C., Erikson, A., et al., 2010, \nat, 464, 384 \\

\bibitem[2009]{Desert2009}D{\'e}sert, J.-M., Lecavelier des 
Etangs, A., H{\'e}brard, G., et al.\ 2009, \apj, 699, 478 

\bibitem[2011a]{Desert11a} 
D\'esert, J.-M., Sing, D. K.,  Vidal-Madjar, A., et al.\ 2011a, \aap, 526, A12

\bibitem[2011b]{Desert11b}D{\'e}sert, J.-M., Bean, J., Miller-Ricci Kempton, E., et al.\ 2011b, \apj, 731, L40 

\bibitem[2016]{Diaz2016}D{\'{\i}}az, R.~F., Rey, J., 
Demangeon, O., et al.\ 2016, \aap, 591, A146 

\bibitem[2007]{Ehrenreich}Ehrenreich, D., H{\'e}brard, G., 
Lecavelier des Etangs, A., et al.\ 2007, \apjl, 668, L179 

\bibitem[2004]{fazio04}
Fazio, G. G., Hora, J. L., Allen, L. E., et al. 2004, \apjs, 154, 10

\bibitem[2002]{han02} 
Han, C., Han, W. 2002, \apj, 580, 490

\bibitem[2010]{Hebrard2010}H{\'e}brard, G., D{\'e}sert, J.-M., 
D{\'{\i}}az, R.~F., et al.\ 2010, \aap, 516, A95 

\bibitem[2015]{Heising2015}Heising, M.~Z., Marcy, G.~W., 
\& Schlichting, H.~E.\ 2015, \apj, 814, 81 

\bibitem[2012]{heller12} 
Heller, R. 2012, \aap, 545, L8

\bibitem[2014]{heller14} 
Heller, R. 2014, \apj, 787, 14

\bibitem[2017]{Heller2017}Heller, R.\ 2017, arXiv:1701.04706 

\bibitem[2014]{Heller2014}Heller, R., Williams, D., Kipping, D., et 
al.\ 2014, Astrobiology, 14, 798 

\bibitem[2010]{hinse10} 
Hinse, T. C., Christou, A. A., Alvarellos, J. L. A., Go\'zdziewski, K. 2010, \mnras, 404, 837

\bibitem[2013]{kalas13} 
Kalas, P., Graham, J. R., Fitzgerald, M. P., Clampin, M. 2013, \apj, 775, 56

\bibitem[2010]{kaltenegger10} 
Kaltenegger, L.\ 2010, \apj, 712, L125

\bibitem[2015]{kenworthy15} 
Kenworthy, M. A., Mamajek, E. E. 2015, \apj, 800, 126

\bibitem[2009a]{kipping09a} 
Kipping, D. M. 2009a, \mnras, 392, 181

\bibitem[2009b]{kipping09b} 
Kipping, D. M. 2009b, \mnras, 396, 1797

\bibitem[2015]{kipping15} 
Kipping, D. M., Schmitt, A. R., Huang, X., et al. 2015, \apj, 813, 14

\bibitem[1993]{laskar93} 
Laskar, J.,Joutel, F., Robutel, P. 1993, \nat, 361, L615

\bibitem[1995]{lecavelier95} 
Lecavelier des Etangs, A., Deleuil, M., Vidal-Madjar, A., et al., 1995, \aap, 299, 557

\bibitem[1999]{Lecavelier99}
Lecavelier des Etangs, A., Vidal-Madjar, A., \& Ferlet, R.\ 1999, \aap, 343, 916 

\bibitem[2016]{lecavelier16} 
Lecavelier des Etangs, A., Vidal-Madjar, A. 2016, \aap, 588, A60

\bibitem[2008]{lewis08} 
Lewis, K. M., Sackett, P. D., Mardling, R. A. 2008, \apj, 685, L153

\bibitem[2013]{Lewis2013}Lewis, K.~M.\ 2013, \mnras, 430, 1473 

\bibitem[2002]{Mandel2002}Mandel, K., 
\& Agol, E.\ 2002, \apjl, 580, L171 

\bibitem[2012]{morbidelli12} 
Morbidelli, A., Tsiganis, K., Batygin, K., Crida, A., Gomes, R. 2012, Icarus, 219, 737

\bibitem[2000]{Nicholson2000}Nicholson, P.~D., French, R.~G., 
Tollestrup, E., et al.\ 2000, \icarus, 145, 474 

\bibitem[2009]{otha09} 
Ohta, Y., Taruya, A., Suto, Y. 2009, \apj, 690, 1

\bibitem[2012]{ogihara12} 
Ogihara, M., Ida, S. 2012, \apj, 753, 60

\bibitem[2012]{Rappaport et al.2012}Rappaport, S., Levine, A., Chiang, E., 
et al.\ 2012, \apj, 752, 1 

\bibitem[2015]{Santos2015}Santos, N.~C., Martins, J.~H.~C., 
Bou{\'e}, G., et al.\ 2015, \aap, 583, A50 

\bibitem[1999]{sartoretti99} 
Sartoretti, P., Schneider, J. 1999, \aaps, 134, 553

\bibitem[1991]{Showalter1991}Showalter, M.~R., Cuzzi, J.~N., 
\& Larson, S.~M.\ 1991, \icarus, 94, 451 

\bibitem[2017]{sicardy2017}
Sicardy, B., El Moutamid, M., Quillen, A. C., et al.\ 2017, in {\it Planetary Ring Systems}, M.~Tiscareno \& C.\ Murray, Eds, Cambridge University Press, arXiv:1612.03321 

\bibitem[2007]{simon07} 
Simon, A. E., Szatm\`ary, K.,  Szab\'o, G. M. 2007, \aap, 470, 727

\bibitem[2010]{simon10} 
Simon, A. E., Szab\'o, G. M., Szatm\`ary, K., Kiss, L. L.  2010, \mnras, 406, 2038

\bibitem[2012]{simon12} 
Simon, A. E., Szab\'o, G. M., Kiss, L. L., Szatm\`ary, K. 2012, \mnras, 419, 164

\bibitem[2010]{Sing2010}Sing, D.~K.\ 2010, \aap, 510, A21 

\bibitem[2012]{Stevenson2012}Stevenson, K.~B., Harrington, J., 
Fortney, J.~J., et al.\ 2012, \apj, 754, 136 

\bibitem[1997]{williams97} 
Williams, D. M., Kasting, J. F., Wade, R. A. 1997, \nat, 385, 234

\bibitem[2015]{Zuluaga2015}Zuluaga, J.~I., Kipping, D.~M., 
Sucerquia, M., \& Alvarado, J.~A.\ 2015, \apjl, 803, L14 

\end{thebibliography}
\end{document}